\documentstyle[12pt,aasms4]{article}
\newbox\grsign \setbox\grsign=\hbox{$>$} \newdimen\grdimen \grdimen=\ht\grsign
\newbox\laxbox \newbox\gaxbox
\setbox\gaxbox=\hbox{\raise.5ex\hbox{$>$}\llap
     {\lower.5ex\hbox{$\sim$}}}\ht1=\grdimen\dp1=0pt
\setbox\laxbox=\hbox{\raise.5ex\hbox{$<$}\llap
{\lower.5ex\hbox{$\sim$}}}\ht2=\grdimen\dp2=0pt

\newcommand{\msun}{$M_\odot$}
\def\simless{\mathbin{\lower 3pt\hbox
   {$\rlap{\raise 5pt\hbox{$\char'074$}}\mathchar"7218$}}} 
\def\simgreat{\mathbin{\lower 3pt\hbox
   {$\rlap{\raise 5pt\hbox{$\char'076$}}\mathchar"7218$}}} 
\def\ltsima{$\; \buildrel < \over \sim \;$}
\def\simlt{\lower.5ex\hbox{\ltsima}}
\def\gtsima{$\; \buildrel > \over \sim \;$}
\def\simgt{\lower.5ex\hbox{\gtsima}}
\def\ref{\noindent\hangindent.5in\hangafter=1}
\def\HI{H{\sc i}~}
\begin{document}
 
\title{STAR FORMATION IN DISK GALAXIES DRIVEN BY PRIMORDIAL H$_2$}

\author{Edvige Corbelli\altaffilmark{1}, Daniele Galli\altaffilmark{1}
and Francesco Palla\altaffilmark{1}}

\altaffiltext{1}{Osservatorio Astrofisico di Arcetri, Largo E.~Fermi 5,
 I--50125 Firenze, Italy}

\authoremail{edvige,galli,palla@arcetri.astro.it}

\begin{abstract}

We show that gaseous \HI disks of primordial composition irradiated by
an external radiation field can develop a multiphase medium with
temperatures between $10^2$ and $10^4$~K due to the formation of
molecular hydrogen.  For a given \HI column density there is a critical
value of the radiation field below which only the cold \HI phase can
exist.  Due to a time decreasing quasar background, the gas starts
cooling slowly after recombination until the lowest stable temperature
in the warm phase is reached at a critical redshift $z=z_{\rm cr}$.
Below this redshift the formation of molecular hydrogen promotes a
rapid transition towards the cold \HI phase. We find that disks of
protogalaxies with $10^{20}\simlt N_{HI}\simlt 10^{21}$ cm$^{-2}$ are
gravitationally stable at $T\sim 10^4$~K and can start their star
formation history only at $z \simlt z_{\rm cr}\sim 2$, after the gas in
the central portion of the disk has cooled to temperatures $T\simlt
300$ K. Such a delayed starbust phase in galaxies of low gas surface
density and low dynamical mass can disrupt the disks and cause them to
fade away.  These objects could contribute significantly to the faint
blue galaxy population.

\end{abstract}

 
\section{Introduction}
  
One of the most interesting and controversial discoveries in the field
of galaxy formation and evolution is related to the excess number count
of Faint Blue Objects (see for example Koo 1996 or Ellis 1997 for a
review).  Various evolutionary models favour a scenario in which
massive galaxies form stars at high redshifts ($z\sim $ 2--3), while
low-mass systems (dwarfs) experience an initial and disruptive
starburst much later and contribute significantly to the Faint Blue
Objects counts (Corbelli $\&$ Salpeter 1995, hereafter CS; Ferguson \&
McGaug 1995; Babul \& Ferguson 1996; Gwyn $\&$ Hartwick 1996; Cowie et
al. 1997; Guzman et al. 1997). The delayed star formation in dwarf
galaxies has been explained as a consequence of the delayed
recombination of ionized hydrogen (Babul \& Rees 1992; Efstathiou 1992)
and of the late onset of the gravitational instability which, in low
mass systems, takes place only when the temperature drops well below
$10^4$ K (Babul \& Rees~1993, CS).

Cooling below $10^4$~K can be caused by a slow build-up of metals, due
to some sporadic star formation events before the onset of the large
scale starburst (CS). However, since the metal abundance required for
efficient cooling increases as the \HI column density of protodisks
decreases, it is likely that additional processes must regulate the
cooling in systems which are at lower end of the gas surface density
distribution.  In this Letter we examine if the formation of molecular
hydrogen  can provide such cooling in gaseous disks.  The role of
molecular coolants has been investigated in various cosmological
contexts, such as the collapse and fragmentation of Jeans unstable
clouds, pregalactic shocks and galaxy formation models (Shapiro \& Kang
1987; Vietri \& Pesce 1995; Ostriker \& Gnedin 1996; Haiman et al.
1995, 1996; Anninos \& Norman~1996; Tegmark et al. 1997; Kepner et
al.~1997).  All these studies have emphasized the importance of
molecular hydrogen as the major agent to drive the temperature of the
gas well below $10^4$~K. The ability of H$_2$ to form in sufficient
amount depends critically on the shielding of the gas to the background
ionizing and dissociating radiation. Here, we study the evolution of
the central regions of a protogalactic disk which at some redshift are
warm and in thermal equilibrium with an external radiation field.  We
show that, as the radiation field decreases with time, the formation of
a small concentration of H$_2$ drives the gas out of equilibrium
causing a rapid transition of the gas from the warm to the cold phase
at a critical redshift $z_{\rm cr}$.  The consequent reduction of
thermal support can cause a gravitational instability  of the disk and
we examine which are the most favorable conditions for this to happen.

\section{Physical Background}

We consider low gas surface density proto-disks (LSPs) embedded in a
spherical dark matter halo and rotationally supported against
gravitational collapse.  Similarly to what is today the disk of a
low-surface brightness galaxy (de Blok et al.~1996), LSPs are expected
to have slowly rising rotation curves, with lower asymptotic values,
$V_{\infty}$, and shallower gas radial distributions compared to disks
with higher gas surface density.  We concentrate on the central regions
of LSPs at $R\simlt R_0$, with $R_0$ in the range 3--8 kpc, where we
can consider the dark matter density and the gas column density as
radially constant.  Here the column density of neutral hydrogen (which
is by far the dominant species) is typically in the range
$10^{20}\simlt N_{HI}\simlt 10^{21}$ cm$^{-2}$, and the corresponding
gas masses inside $R_0$ are between $10^7$--$10^9$~\msun. The dark
matter determines the value of $V_{\infty}$; for
$V_{\infty}=30$--150~km~s$^{-1}$ the total mass inside $R_0$ is
$10^8$--$10^{10} $~\msun. Disks with such characteristics are expected
to exist already at $z\simeq 2$, as shown by the theoretical
calculations by Kauffmann~(1996) and by observations of damped
Ly$\alpha$ systems (Wolfe et al.~1994).

After recombination the vertical gas distribution is nearly
isothermal, in hydrostatic equilibrium at each radius. 
It will be shown in the next Section that the gas 
stays quite uniform in temperature for most of the time and therefore  
in this Letter we do not consider explicitly its vertical stratification. 
We approximate  the total gas pressure with its value at midplane as 
$$
{P\over k} \simeq 
40\Bigl({N_{HI}\over 10^{20}{\hbox{cm}}^{-2}}\Bigr)^2
+ 2 \Bigl({N_{HI}\over 10^{20}{\hbox{cm}}^{-2}}\Bigr)
\Bigl({c_s\over {\hbox{km s}}^{-1}}\Bigr)
\Bigl({V_{\infty}\over {\hbox{km s}}^{-1}}\Bigr)
\Bigl({R_0\over {\hbox{kpc}}}\Bigr)^{-1}
\qquad {\hbox{cm}}^{-3} ~{\hbox{K}}
\eqno (2.1)
$$
where $k$ is the Boltzmann constant, and $c_s$ is the isothermal sound
speed. The two terms in eq. (2.1) represent the contributions of the
gas self-gravity and the dark matter, respectively.  The pressure is
radially constant in the central region and changes with time as the
slab cools due to the temperature dependence of $c_s$, which we follow
explicitly.

The chemistry of H$_2$ depends upon the density of free electrons and
therefore on the ionizing flux at energies $E>100$~eV, able to
penetrate \HI column densities $\simgt 10^{20}$~cm$^{-2}$, as well as
on the UV flux below the Lyman limit. At redshifts $z\simlt 2$, the
quasar background decreases with time and accounts for most of the
ionizing flux (Irwin, McMahon, \& Hazard 1991).  We shall use the
following fit to the results of Haardt \& Madau~(1996) at $z=1$
$$
J(E,z)= \Bigl({1+ z\over 2}\Bigr)^3 \times  \cases
{A\times 1.7\times 10^{-21} \Bigl({E \over 4.96~} \Bigr)^{-0.3}
&for $E<$ 4.96 \cr
A\times 1.5\times 10^{-22} \Bigl({E \over 13.6~} \Bigr)^{-2.4}
&for 4.96$\le E < $13.6 \cr
B\times  5.0\times 10^{-24} \Bigl({E \over 13.6~} \Bigr)^{-0.3}
&for 13.6$\le E <$ 413 \cr
B\times  1.8\times 10^{-24} \Bigl({E \over 413~} \Bigr)^{-1.5}
&for $E \ge $413 \cr}
\eqno (2.2)
$$
where $E$ is in eV, and $J(E,z)$ in
erg~s$^{-1}$~cm$^{-2}$~sr$^{-1}$~Hz$^{-1}$. For the Haardt \&
Madau~(1996) spectrum, $A=B=1$ which defines our {\it standard} flux
spectrum.  We consider variations of $A$ and $B$ to account for
uncertainties in the intrinsic emission spectrum of quasars, in the
attenuation by intervening clouds, and in additional UV radiation from
nearby bright star-forming galaxies or from sporadic internal star
formation.

We have followed the evolution of $e$, H, H$^+$, H$^-$ and H$_2$ by
means of a reduced chemical network which includes the most relevant
reactions listed in Table~1.  We have also considered the H$_2^+$
channel for H$_2$ formation, but its contribution was found to be
negligible.  Since temperatures of interest in this paper are below
$10^4$~K, secondary electrons from H and He have been included
following the prescription of Shull \& van Steenberg (1985).  These
electrons tend to level the difference between He and H fractional
ionization deriving from direct photoionization alone.  Thus, the
residual abundance of He$^+$ is set equal to that of H$^+$.  For the
photoionization rates we have integrated the cross sections given in
the references listed in Table 1 over the background spectrum.  We
account for the attenuation of the flux as it penetrates the slab to
the midplane assuming that the density of the chemical species is
uniform.  The recombination coefficient for hydrogen includes all the
recombinations to $n\ge 2$ (on the spot approximation) and we assume
that all the recombinations of He produce one ionizing photon for H.
The line-shielding of H$_2$ molecules in the ground level has been
computed using the prescriptions of Federman et al. (1979) in the
absence of dust grains. The photodissociation probabilities and
oscillator strengths for Lyman and Werner bands are from Dalgarno \&
Stephens (1970) and Abgrall \& Roueff (1989).

The thermal evolution of the gas is computed by solving the energy
equation together with the chemical network (we have used a Hubble
constant of 75 km s$^{-1}$ Mpc$^{-1}$). Heating of the gas is provided
by photoionization of atomic H and He (Shull $\&$ van Steenberg 1985).
Cooling is due to electron-impact excitation of atomic H (Spitzer 1978)
and to H-impact excitation of roto-vibrational levels of H$_2$
(Hollenbach \& McKee 1989, but with the correct sign as in Flower et
al. 1986).  We have neglected the contribution of H$_2$-H$_2$
collisions to the cooling because of the small molecular abundance.

\section{The two phase equilibrium and the fast gas cooling}

For a given value of $V_{\infty}/R_0$ the gas temperature of a slab in
thermal equilibrium  depends on the intensity of the background field
and on the gas column density of the slab. For the standard background
spectrum Fig.~1a (curves 1 and 2) shows the possible equilibrium
temperatures, after the gas has recombined, as function of redshift for
two representative values of $N_{HI}$.

We see from Fig.~1a that conditions for a multiphase medium, in which
gas at different temperatures coexists in pressure balance, can be
achieved over a significant range of redshift. This is analogous to
what happens in the present-day interstellar medium (Field et. al
1969), where metal lines radiation rather than $H_2$ dominates the
cooling.  In general a multiphase medium is  possible if for a given
pressure the cooling per atom is a non monotonic function of $T$. In
the present context at $T\simgt 7000$ K the cooling rate is an
increasing function of $T$ since the dominant process is the e-H impact
excitation. Below 7000~K, $H-H_2$ impact excitation becomes the
relevant cooling process:  H$_2$ is formed efficiently by reaction (3)
and is destroyed by reaction (5). The strong temperature dependence of
reaction (5), proportional to $e^{-21300/T}$, allows a rapid rise of
the H$_2$ fractional abundance as the temperature diminishes. Due to
the weak temperature dependence of the $H_2$ cooling function in this
region, cooling per particle becomes a decreasing function of $T$ and
the equilibrium curve has an inflection. This is a thermally unstable
regime which ends at $T\simlt 2000$~K,  when the 2-step
photodissociation takes over reaction (5) (direct dissociation by hard
photons, reaction 8, never contributes significantly to the H$_2$
balance).  Under some circumstances it may not be obvious to determine
the dominat thermal phase if, for a given ionizing flux, three
equilibrium temperatures are possible. For the warm slab considered
here the temperature evolution is much simpler: as the background flux
declines with time, the temperature decreases monotonically from $\sim
10^4$~K moving along the equilibrium curve towards the right hand side
of Fig..~1a.  The non monotonic relation $T(z$) implies that at a
certain redshift $z_{\rm cr}$ the gas reaches the lowest stable
equilibrium temperature in the warm phase: we shall call this {\it
transition point} (marked by an asterisk in Fig. 1).  For a slab of
typical column density $N_{HI}=10^{20.4}$ cm$^{-2}$  $z_{\rm cr}$ is
$\simeq 0.5$. The gas is thereafter forced to get out of equilibrium. 
The dashed tracks of Figure 1a show the subsequent time evolution of $T$
computed for a parcel of gas close to the midplane. We can see that the
gas cools quickly toward the cold stable phase spending a very short
interval of time at temperatures between 7000~K and 200~K.  The
transition towards the cold phase propagates rapidly vertically since
the upper layers loose thermal support and settle towards the center
incresing their density and cooling further. Due to the rapid cooling,
shortly after the cold core forms most of the gas will be at close
temperatures and the warm atmosphere left above contains little mass.
During the transition the fractional abundance of H$_2$, $f$(H$_2$),
increases from $\sim 3 \times10^{-5}$ to $\sim 10^{-3}$ and then levels
off (Fig.~1b).  The final value of $f$(H$_2$) is not very sensitive to
either the background flux or the gas column density.

LSPs with $N_{HI}$ column density between 10$^{20.1}$ and 10$^{20.8}$
cm$^{-2}$ make a transition to the cold phase in the redshift interval
$z=2$ to $z=0$.  The value of $z_{\rm cr}$ can vary appreciably with
the intensity of the background flux. From curve 3 of Fig.~1$(a)$ we
can see in fact that  as $A$ increases, cooling of the slab is delayed
until a much lower $z$.  We have examined cases with $0.1<A<10$ and
$0.3<B<3$ and the important point is that there is always a range of
column densities between $\sim 10^{20}$ and $\sim 10^{21}$ cm$^{-2}$
which make the transition to the cold phase in the redshift interval
$z=0$--2.  This conclusion holds even if a small amount of metals (up
to 0.01 solar) is present in the gas mixture as a contamination from an
early generation of stars. Therefore, for the central regions of LSPs,
molecular hydrogen plays a fundamental role in promoting and driving
the thermal evolution of the gas.

\section{Gravitational instability and the star formation phase}

Protodisks exposed to a time-varying external radiation
field follow the evolution described in the previous section if they are
still gaseous and gravitationally stable in the warm phase.  We examine
the global gravitational stability of the simple hydrostatic model
described in Sect.~2 using the Toomre criterion (Toomre~1964; see also CS).
For $R<R_0$ the parameter $Q$ can be written as
$$
Q\simeq 0.5  \Bigl({T\over 7000{\hbox{K}}}\Bigr)^{1/2}
\Biggl\lbrack\Bigl({N_{HI}\over 10^{20}{\hbox{cm}}^{-2}}\Bigr)^{-2}
\Bigl({V_{\infty}\over {\hbox{km s}}^{-1}}\Bigr)^2
\Bigl({R_0\over {\hbox{kpc}}}\Bigr)^{-2}+
12\Bigl({N_{HI}\over 10^{20}{\hbox{cm}}^{-2}}\Bigr)^{-1}
\Biggr\rbrack^{1/2}\eqno (4.1)
$$
To evaluate the contribution of the self-gravity term we have used eq.
(25) of Toomre (1963) assuming that the protogalaxy has a total gas
mass $M_g\simeq 4\times 10^8 (N_{HI}/ 10^{20})$ M$_\odot$.  This
guarantees a radially constant gas surface density for $R<R_0$ and a
slowly rising rotation curve with $V_\infty/R_0$ determined by the
amount of dark matter. Therefore, for any $N_{HI}$, we can derive  $Q$
as function of redshift.  We are interested in all the disk models
which are stable in the warm phase but become unstable ($Q<1$) only at
$z<z_{\rm cr}$ due to the decrease of the isothermal sound speed driven
by H$_2$ cooling.  These galaxies will experience a delay in the onset
of large scale star formation.

A comprehensive view of the stability and dominant thermal phases of
the gas in the plane $N_{HI}-V_{\infty}/R_0$ for the standard
background spectrum is given in Figure~2.  In the region labelled {\it
cold $\&$ unstable} slabs which are stable for $z<z_{\rm cr}$ become
gravitationally unstable before $z=0$.  LSPs in this region span a
considerable range of \HI column densities below $10^{21}$ cm$^{-2}$.
Models in the {\it warm $\&$ unstable} region have $Q<1$ for $T=7000$~K
and the collapse starts as soon as the gas settles into the disk.  The
shaded regions in the left part of Fig.~2 show stable models which can
be either cold or warm but cannot form stars.  For some of these slabs
cooling of the gas did actually start, but the temperature was not low
enough to trigger the instability before $z=0$ ({\it cold $\&$
stable}). Since cold \HI disks have not been observed, it is likely
that LSPs were never formed with these initial conditions (very low
$N_{HI}$ and high $V_{\infty}/R_0$ values).  Finally, in the {\it warm
$\&$ stable} region the gas remains warm up to $z=0$; here the
parameters of the slabs are more appropriate for the outer regions of
disk galaxies (Corbelli \& Salpeter 1993).  Outer regions of LSPs and
their central regions prior to $z_{\rm cr}$ might also contribute
significantly to QSO absorption lines (see also Linder 1997) because
the time the gas spends in the warm phase at $T \simgt 7000$~K is long
compared to the transition time toward the cold and disruptive phase.

The dotted lines in the {\it cold $\&$ unstable} region of Fig.~2
represent the loci of $Q=1$ and are labelled with the corresponding
redshift (the $z=0$ line coincides with the heavy solid line at the
left hand side of this region).  Systems of increasingly lower \HI
column density start to cool and become unstable at lower and lower
redshifts. For values of $A$ and $B$ higher than the standard case, the
left border of the {\it cold $\&$ unstable} region and the dotted lines
shift to the right.  If at some redshift slabs of higher $N_{HI}$ had
already become cold and unstable but the background flux intensity
receives an extra input, the whole cooling process might stop.  This
can explain the lack of a large population of faint star forming disks
of very low column density at $z=0$. On the other hand, the observed
deficit can also be due to a truncation of the distribution function of
protodisks of low column densities.

How does the theoretical scheme of Fig.~2 compare with observations?  A
precise answer to this question is difficult since the parameters
$N_{HI}$ and $V_{\infty}/R_0$ of protogalactic disks are unknown.
However, as an illustration, we can use the results of de Blok et al.
(1996) on Low Surface Brightness galaxies (LSBs) which are reminiscent
of our models of LSPs.  We have estimated $N_{HI}$ and $V_{\infty}/R_0$
for ten galaxies of their sample with slowly rising rotation curves,
excluding LSBs with steeply declining gas column density profiles in
the inner regions.  The selected objects correspond to LSBs with
$10<R_{25}<30$ arcsec and the results are shown by the filled triangles
of Fig.~2. The data fall within the boundaries of the {\it cold $\&$
unstable} region and indicate that if LSPs were similar to LSBs they
had the adequate parameters to follow the evolution described in this
Letter. Within this region, systems of low column density and dynamical
mass are those for which the ensuing starburst easily disrupts the
gaseous central region and reduces considerably the gas mass content.
Thus, further star formation is inhibited and the systems experience a
fading away phase (CS).  The identification of LSPs with the
progenitors of LSBs is quite appealing since some of the properties of
the blue galaxies observed at moderate $z$ can be explained if these
are edge-on manifestations of LSBs (Dalcanton \& Shectman 1996).
Present-day LSBs might descend from those LSPs which, for environmental
effects or some internal feature (higher dark matter or gas content),
had a  slower, and less violent star formation phase.

We are grateful to E. E. Salpeter and to the referee for their comments
and to R. Baglioni for his help with the figures. This work was
supported in part by ASI-95-RS-120.

\clearpage

\begin{table*}
\caption[]{List of reactions} 
\begin{flushleft}
\begin{tabular}{cll} \hline \hline
 \#  & Reaction & Reference  \\ \hline
1 &H$^+$ $+$ $e$ $\rightarrow$ H $+$ $h\nu$ &  Seaton 1959 \\
2 &H $+$ $e$ $\rightarrow$ H$^-$ $+$ $h\nu$ & De Jong 1972 \\
3 &H$^-$ $+$ H $\rightarrow$ H$_2$ $+$ $e$ & Bieniek 1980 \\
4 &H$^-$ $+$ H$^+$ $\rightarrow$ 2H & Peterson et al. 1971 \\
5 &H$_2$ $+$ H$^+$ $\rightarrow$ H$_2^+$ $+$ H & Hollenbach \& McKee 1989 \\
6 &H $+$ $h\nu$ $\rightarrow$ H$^+$ $+$ $e$ & Cen 1992  \\
7 &H$^-$ $+$ $h\nu$ $\rightarrow$ H $+$ $e$ & Wishart 1979 \\
8 &H$_2$ $+$ $h\nu$ $\rightarrow$ H$_2^+$ $+$ $e$ & O'Neil \& Reinhardt 1978\\
9 &H$_2$ $+$ $h\nu$ $\rightarrow$ 2H & see text \\
\hline

\end{tabular}
\end{flushleft}
\end{table*}

\clearpage

\begin{center}
{\bf FIGURE CAPTIONS}

\figcaption[]{
$(a)$ Equilibrium gas temperatures as function of redshift for three
cases: (1)~$N_{HI}=10^{20.6}$ cm$^{-2}$ and $A=B=1$,
(2)~$N_{HI}=10^{20.4}$~cm$^{-2}$ and $A=B=1$, and
(3)~$N_{HI}=10^{20.4}$~cm$^{-2}$ and $A=10,B=1$ (continuos lines). The
dashed lines show the time evolution of the temperature once the slab
reaches the transition point marked with the asterisk.  $(b)$ 
Fractional abundances of H$_2$ after the transition
point and prior to $z=0$ for the same cases as in $(a)$.  For all cases
shown $V_\infty/R_0 =15$ km s$^{-1}$ kpc$^{-1}$.}

\figcaption[]{Regions in the N$_{HI}$--$V_{\infty}/R_0$ plane
corresponding to the different thermal and dynamical regimes discussed
in the text. The units of $V_{\infty}/R_0$ are km s$^{-1}$ kpc$^{-1}$,
$N_{HI}$ is in cm$^{-2}$. The dotted lines connect points which become
gravitationally unstable at the same redshift and are labelled
accordingly.}

\end{center}

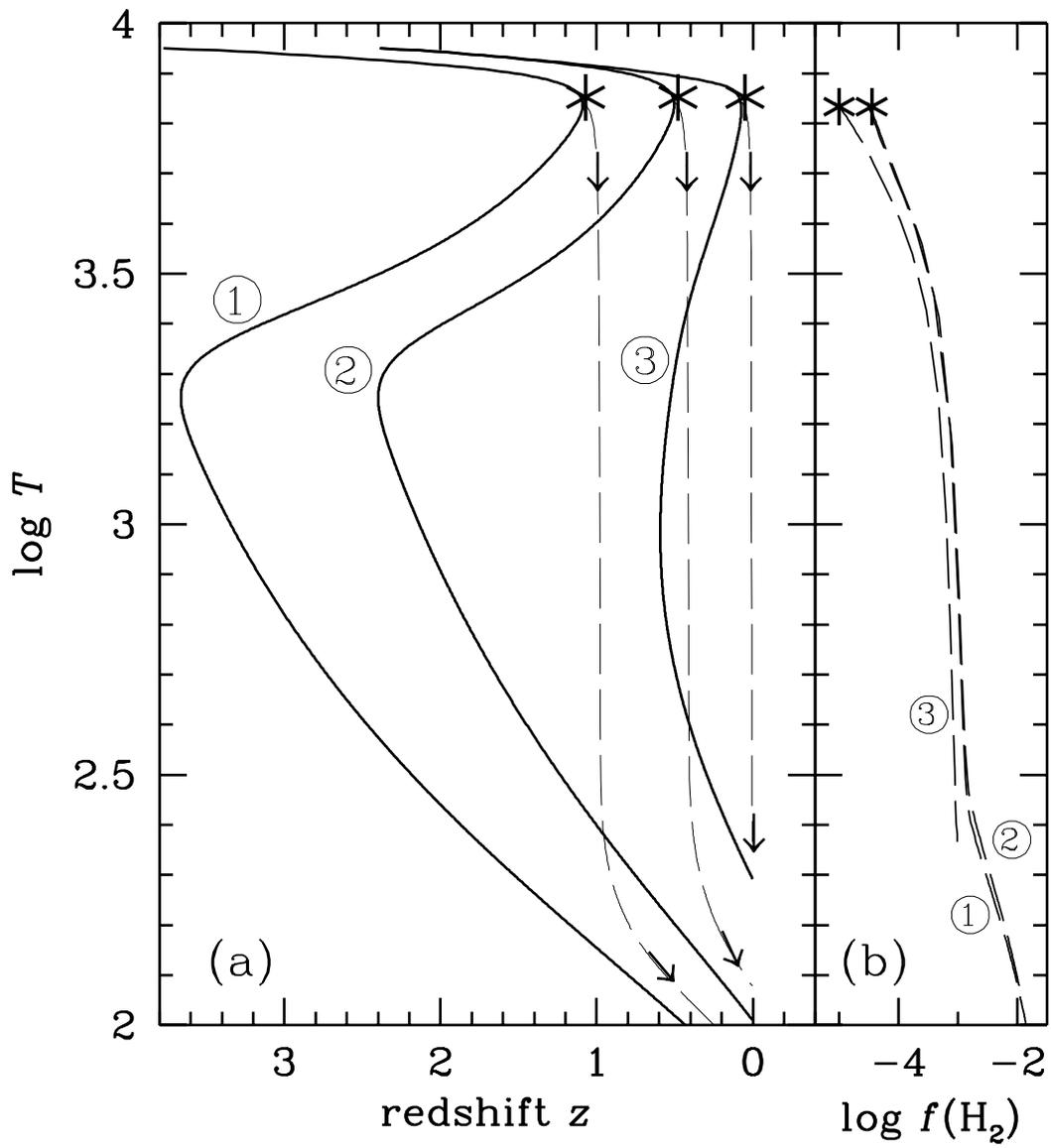
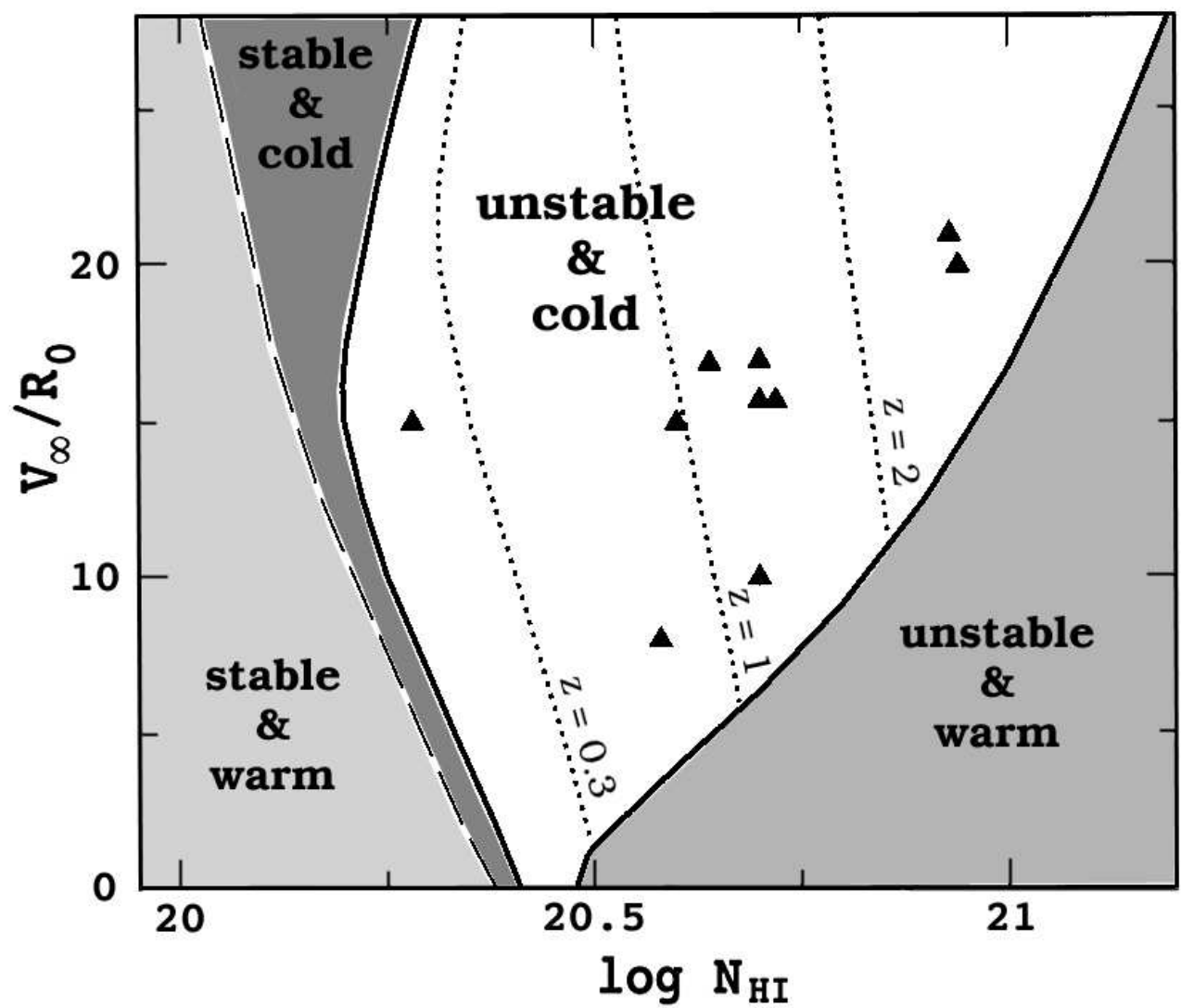
\end{document}